\documentclass[twocolumn,aps,prl,showpacs]{revtex4}
\usepackage[latin9]{inputenc}
\usepackage{amsmath}
\usepackage{amssymb}
\usepackage{graphicx}
\usepackage{esint}

\makeatletter
\@ifundefined{textcolor}{}
{%
 \definecolor{BLACK}{gray}{0}
 \definecolor{WHITE}{gray}{1}
 \definecolor{RED}{rgb}{1,0,0}
 \definecolor{GREEN}{rgb}{0,1,0}
 \definecolor{BLUE}{rgb}{0,0,1}
 \definecolor{CYAN}{cmyk}{1,0,0,0}
 \definecolor{MAGENTA}{cmyk}{0,1,0,0}
 \definecolor{YELLOW}{cmyk}{0,0,1,0}
}

\makeatother

\begin{document}

\title{Spin-orbit-coupled topological Fulde-Ferrell states of fermions in
a harmonic trap}

\author{Lei Jiang$^{1}$, Eite Tiesinga$^{1}$, Xia-Ji Liu$^{2}$, Hui Hu$^{2}$
and Han Pu$^{3}$}

\affiliation{$^{1}$Joint Quantum Institute, University of Maryland and National
Institute of Standards and Technology, Gaithersburg, Maryland 20899,
USA\\
 $^{2}$Centre for Atom Optics and Ultrafast Spectroscopy, Swinburne
University of Technology, Melbourne 3122, Australia\\
 $^{3}$Department of Physics and Astronomy, and Rice Quantum Institute,
Rice University, Houston, Texas 77251, USA}
\begin{abstract}
Motivated by recent experimental breakthroughs in generating spin-orbit
coupling in ultracold Fermi gases using Raman laser beams, we present
a systematic study of spin-orbit-coupled Fermi gases confined in a
quasi-one-dimensional trap in the presence of an in-plane Zeeman field
(which can be realized using a finite two-photon Raman detuning).
We find that a topological Fulde-Ferrell state will emerge, featuring
finite-momentum Cooper pairing and zero-energy Majorana excitations
localized near the edge of the trap based on the self-consistent Bogoliubov-de
Genes (BdG) equations. We find analytically the wavefunctions of the
Majorana modes. Finally using the time-dependent BdG we show how the
finite-momentum pairing field manifests itself in the expansion dynamics
of the atomic cloud.
\end{abstract}

\date{\today }


\maketitle
Over the past few years, spin-orbit coupled quantum gases have received
a great amount of interest in both the cold atom and condensed matter
communities \cite{sau10,Oreg10,Mourik12,lin11,sato09,shenoy11,hu11,zhai11,chuanwei11,yi11,iskin11,melo12}.
This can be largely attributed to the fact that such a system can
potentially realize exotic quantum phases in a controlled fashion.
So far only one type of spin-orbit coupling (SOC) --- the equal-weight
Rashba and Dresselhaus SOC --- has been realized \cite{wang12,cheuk12},
although several theoretical schemes have been proposed to realize
more general types of SOC. Nevertheless, the experimentally realized
SOC has already been shown to give rise to several interesting quantum
phases. These include the topological superfluid phase in a one-dimensional
(1D) system supporting Majorana modes near the boundaries \cite{liu12a,liu12b,wei12,mizushima13},
and the Fulde-Ferrell (FF) superfluid state featuring finite-momentum
Cooper pairing \cite{dong13a,dong13b,shenoy13,Zheng13,wu13,liu13a,liu13b}.
Furthermore, in a 1D setting, these two features can coexist where
one realizes an exotic topological FF superfluid \cite{qu13,zhang13,liu13c,chen13,ruhman13}.
Early experiments explored the possibility of the Fulde-Ferrell-Larkin-Ovchinnikov
state in spin-imbalanced cold atoms \cite{Zwierlein06,Partridge06,Nascimbene09},
with indirect evidence only coming from a quasi-1D setup \cite{Liao10}.

Previous work on topological FF phases focused on homogeneous 1D gases.
In this work, we consider a system in a realistic 1D harmonic trapping
potential and investigate how signatures of exotic phases can be probed
in practice. Such a quasi-1D system can be realized by confining atoms
in strong 2D optical lattices \cite{Liao10}. The presence of a trapping
potential suppresses quantum fluctuations and makes mean-field calculations
qualitatively reliable \cite{Sundar13,liu07}. The two goals of this
work are: (1) We show how a Majorana mode, the smoking gun of the
topological order, manifests itself in the density of states (DOS)
in both real and momentum space. We obtain the wavefunctions of the
Majorana quasi-particle states analytically and show that they are
in good agreement with numerical results based on the Bogoliubov-de
Gennes (BdG) formalism. (2) We show how finite-momentum Cooper pairing,
the telltale signal of the FF phase, leaves detectable traces in the
expansion dynamics of the atomic cloud.

{\em Model ---} We consider a spin-1/2 Fermi gas confined in a
1D harmonic trap. Its Hamiltonian is given by $H=H_{0}+H_{{\rm int}}$,
with single-particle component $H_{0}$ and interacting component
$H_{{\rm int}}$ describing the $s$-wave contact interaction. In
fact \begin{widetext}
\begin{eqnarray*}
H_{0} & = & \int dx\,\psi_{\uparrow}^{\dag}(x)[H_{s}+\delta/2]\psi_{\uparrow}(x)+\int dx\,\psi_{\downarrow}^{\dag}(x)[H_{s}-\delta/2]\psi_{\downarrow}(x)-\frac{\Omega_{R}}{2}\int dx\,\left[\psi_{\uparrow}^{\dag}(x)e^{i2k_{R}x}\psi_{\downarrow}(x)+H.c.\right]\\
H_{{\rm int}} & = & g_{1D}\int dx\,\psi_{\uparrow}^{\dag}(x)\psi_{\downarrow}^{\dag}(x)\psi_{\downarrow}(x)\psi_{\uparrow}(x),
\end{eqnarray*}
 \end{widetext} where $\psi_{\uparrow}(x)$, $\psi_{\downarrow}(x)$
are Fermi annihilation operators for the two spin states. The Hamiltonian
$H_{s}=-\frac{\hbar^{2}}{2m}\frac{\partial^{2}}{\partial x^{2}}-\mu+V_{T}(x)$
with $\mu$ being the chemical potential and $V_{T}(x)=\frac{1}{2}m\omega^{2}x^{2}$
the harmonic trapping potential with a frequency $\omega$. The constants
$\delta$ and $\Omega_{R}$ represent the detuning and strength of
the two-photon Raman coupling, respectively, $2\hbar k_{R}$ is the
photon recoil momentum imparted to the atoms from the Raman lasers,
and $\hbar$ is reduced Planck constant. Finally, $g_{1D}$ is the
1D two-body $s$-wave interaction strength.

After applying the local gauge transformation
\begin{eqnarray*}
\psi_{\uparrow}(x) & = & e^{ik_{R}x}\,[\phi_{\uparrow}(x)-i\phi_{\downarrow}(x)]/\sqrt{2}\,,\\
\psi_{\downarrow}(x) & = & e^{-ik_{R}x}\,[\phi_{\uparrow}(x)+i\phi_{\downarrow}(x)]/\sqrt{2}\,,
\end{eqnarray*}
 the single-particle Hamiltonian $H_{0}$ becomes
\begin{equation}
H_{0}=\int dx\,\phi^{\dag}[H_{s}+(-i\lambda\partial_{x}+\nu)\sigma_{y}-h\sigma_{z}]\,\phi\,,\label{hs}
\end{equation}
 where $\phi=[\phi_{\uparrow}(x),\phi_{\downarrow}(x)]^{T}$ and we
have dropped a constant corresponding to the atomic recoil energy.
In writing Eq.~(\ref{hs}), we have defined the spin-orbit-coupling
constant $\lambda\equiv\hbar^{2}k_{R}/m$, the effective out-of-plane
Zeeman field $h\equiv{\Omega_{R}}/{2}$ and the effective in-plane
Zeeman field $\nu\equiv\delta/2$. It is convenient to define effective
Zeeman field strength $b\equiv\sqrt{h^{2}+\nu^{2}}$. The operators
$\sigma_{y}$ and $\sigma_{z}$ are Pauli matrices in the atomic spin
basis. The interaction Hamiltonian $H_{{\rm int}}$ is invariant under
this gauge transformation.

{\em Bogoliubov-de~Gennes formalism ---} In the mean-field BdG
approximation, we assume a non-zero complex order parameter or gap
$\Delta(x)\equiv-g_{1D}\langle\phi_{\downarrow}(x)\phi_{\uparrow}(x)\rangle=-ig_{1D}\langle\psi_{\downarrow}(x)\psi_{\uparrow}(x)\rangle$,
and in terms of the Nambu spinor $\Phi(x)=[\phi_{\uparrow}(x),\phi_{\downarrow}(x),\phi_{\uparrow}^{+}(x),\phi_{\downarrow}^{+}(x)]^{T}$
the mean-field Hamiltonian becomes
\[
H_{{\rm mf}}=\frac{1}{2}\int dx\,\Phi^{\dag}(x)H_{{\rm BdG}}\Phi(x)+{\rm Tr}[H_{s}]-\int dx\,\frac{|\Delta(x)|^{2}}{g_{1D}}\,,
\]
 where
\[
H_{{\rm BdG}}=\left[\begin{array}{cccc}
H_{s}-h & -\lambda\partial_{x}-i\nu & 0 & -\Delta(x)\\
\lambda\partial_{x}+i\nu & H_{s}+h & \Delta(x) & 0\\
0 & \Delta^{\ast}(x) & -H_{s}+h & \lambda\partial_{x}-i\nu\\
-\Delta^{\ast}(x) & 0 & -\lambda\partial_{x}+i\nu & -H_{s}-h
\end{array}\right]\,.
\]

The Bogoliubov quasi-particles are obtained by diagonalizing
\begin{equation}
H_{{\rm BdG}}\,\varphi_{\eta}(x)=E_{\eta}\,\varphi_{\eta}(x)\,,\label{bdg}
\end{equation}
 with energies $E_{\eta}$ and wavefunctions $\varphi_{\eta}(x)=[u_{\uparrow\eta}(x),u_{\downarrow\eta}(x),v_{\uparrow\eta}(x),v_{\downarrow\eta}(x)]^{T}$
 indexed by subscript $\eta=1,2,3\ldots$
The wavefunctions are normalized such that ${\textstyle \sum_{\sigma=\uparrow,\downarrow}}\int dx(|u_{\sigma\eta}(x)|^{2}+|v_{\sigma\eta}(x)|^{2})=1.$
The order parameter
\[
\Delta(x)=-\frac{g_{1D}}{2}\underset{\eta}{\sum}\,[u_{\uparrow\eta}v_{\downarrow\eta}^{\ast}f(E_{\eta})+u_{\downarrow\eta}v_{\uparrow\eta}^{\ast}f(-E_{\eta})]\,
\]
must be solved self consistently, where $f(E)$ is Fermi-Dirac distribution
function $f(E)=1/[e^{E/k_{B}T}+1]$ and $T$ is the temperature. Here
we present results for $T=0$.

To solve the eigenvalue problem, we use the discrete variable representation
of the plane wave basis \cite{colbert92}. We employ $1001$ plane-wave
bases and the total number of atoms is $N=60$. In the harmonic trap
with frequency $\omega$, we define the non-interacting Fermi energy
$E_{F}=\hbar\omega{N}/{2}$, the Fermi wave number $k_{F}$, obtained
from $E_{F}=\hbar^{2}k_{F}^{2}/(2m)$, and the Thomas-Fermi radius
$x_{{\rm TF}}=\sqrt{N\hbar/(m\omega)}$. Throughout we use $E_{F}$
and $x_{TF}$ as the natural energy and length scale, respectively.
Equation (\ref{bdg}) is solved by using a ``hybrid'' method of
Ref. \cite{liu07,liu13d}. We start with an initial guess of the order
parameter. We then find all the eigenpairs of $H_{\rm BdG}$ with energy
$|E|\leqslant E_{c}$ where $E_{c}$ is a cut-off energy that is chosen
to be large compared to the Fermi energy but small compared to the
full spectral width of the discretized $H_{\rm BdG}$. Typically we take
$E_{c}=8E_{F}$. For states above the energy cut-off, we employ a
semi-classical method based on the local density approximation. The
updated order parameter is calculated by combining the contributions
from the numerical and semi-classical solutions. The procedure is
repeated until convergence of the order parameter is reached.

It is convenient to characterize the interaction strength $g_{1D}$
by a dimensionless interaction parameter $\gamma\equiv-mg_{1D}/(\hbar^{2}n_{0})$
which represents the ratio between the interaction and kinetic energy.
Here $n_{0}=(2/\pi)\sqrt{Nm\omega/\hbar}$ is the total atomic density
at the trap center in the Thomas-Fermi approximation.

To study expansion dynamics, we solve the time-dependent BdG equation
\cite{tonini06,challis07,scott11,lu12}
\[
i\hbar{\partial_{t}}\,\varphi_{\eta}(x,t)=H_{{\rm BdG}}\,\varphi_{\eta}(x,t)\,,
\]
 with the initial wavefunction of $\varphi_{\eta}(x,0)$ set by the
ground state of the trapped system obtained from Eq.~(\ref{bdg}).
We use the Runge\textendash{}Kutta method for the time propagation.
The kinetic energy and the spin-orbit coupling term are propagated
in the interaction picture using a fast Fourier transform.

\begin{figure}
\includegraphics[width=0.45\textwidth]{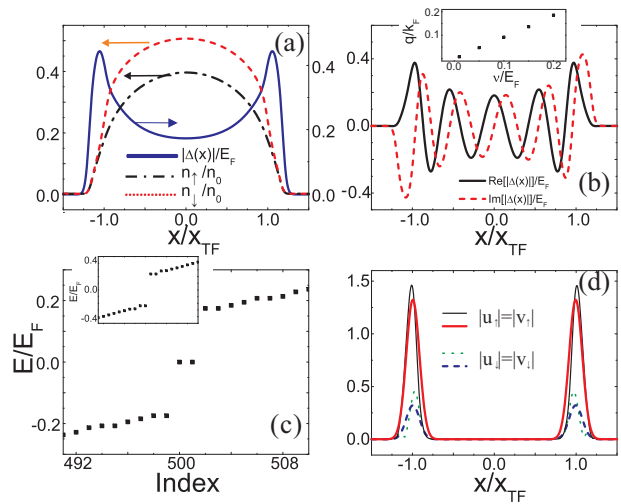} \caption{(a) Density profiles $n_{\sigma}(x)$ in units of $n_{0}$ (left y-axis)
and order parameter profile $|\Delta(x)|$ in units of $E_{F}$ (right
y-axis). (b) The real and imaginary parts of the order parameter profile.
The inset shows the pairing momentum $q$ as a function of $\nu$.
(c) Low-lying spectrum of quasi-particle excitation. The inset shows
the corresponding spectrum of a topologically trivial phase with $\nu=0$
and $h=0.5E_{F}$ with other parameters the same as those of the rest
of the figure. (d) Wavefunction of the Majorana modes. The thin (thick)
lines represent numerical (analytical) results. The parameters used
for this figure are $\gamma=2$.2, $\lambda=1.5E_{F}/k_{F}$, $h=0.8E_{F}$,
and $\nu=0.2E_{F}$ unless otherwise noted.}

\label{f1}
\end{figure}

{\em Topological FF state in trap ---} In Fig.~\ref{f1}, we characterize
the properties of the topological FF state in the trapped system using
experimentally relevant parameters. Figure~\ref{f1}(a) shows atom
density profiles of both spin species, $n_{\sigma}(x)=\langle\psi_{\sigma}^{\dag}(x)\psi_{\sigma}(x)\rangle$,
along with the magnitude of the order parameter $|\Delta(x)|$. The
densities peak at the center of the trap, while $|\Delta(x)|$ has
a minimum at the center and reaches its maximum value near the edge
of the cloud. This is due to a peculiar property of 1D quantum gases
\cite{liu07}: for sufficiently large density, the effect of the interaction
and thus the order parameter is enhanced by reducing the number density,
while near the edge of the cloud, the order parameter has a power
law dependence on density.

For positive two-photon detuning $\delta=2\nu$ used in the calculation,
the effective chemical potential for spin-up atoms, $\mu-\delta/2$,
is lower than that for spin-down atoms, $\mu+\delta/2$ . Consequently,
spin-up (spin-down) represents the minority (majority) species. As
we change the sign of $\delta$, the density profiles of the two spin
species switch. This follows from a symmetry of the BdG Hamiltonian
for this system: Under the simultaneous transformation $\nu\rightarrow-\nu$, $\Delta(x)\rightarrow-\Delta(-x)$
and 
\begin{eqnarray*}
 & [u_{\uparrow\eta}(x),\, u_{\downarrow\eta}(x),\, v_{\uparrow\eta}(x),\, v_{\downarrow\eta}(x)]\rightarrow\\
 & \;\;\;\;\;\;\;\;\;\;\;\;\;[u_{\uparrow\eta}(-x),\,-u_{\downarrow\eta}(-x),\, v_{\uparrow\eta}(-x),\,-v_{\downarrow\eta}(-x)]\,,
\end{eqnarray*}
Equation~(\ref{bdg}) remains invariant.

Figure~\ref{f1}(b) shows the real and imaginary parts of the order
parameter. Near the center of the trap, the order parameter has the
plane-wave form $\Delta(x)\simeq\Delta_{0}\, e^{iqx}$, indicating
finite-momentum pairing. This is the characteristic signature of a
FF superfluid state \cite{fulde64,larkin65}. The pairing momentum
$q$ can be easily extracted from the data. The inset of Fig.~\ref{f1}(b)
shows that $q$ increases nearly linearly as a function of $\nu$.

{\em Majorana modes ---} In Fig.~\ref{f1}(c), we show the low-lying
quasi-particle excitation spectrum of the system. The two dots with
zero energy are the Majorana modes characterizing the nontrivial topological
nature of the system. The Majorana modes exist inside a gap in the
energy spectrum. For comparison, a spectrum for a topologically trivial
system is shown in the inset, where no states inside the gap exist.
The wavefunctions of the two Majorana modes are plotted in Fig.~\ref{f1}(d).
The two Majorana modes are spatially localized near the two edges
of the cloud at $x=\pm x_{TF}$. Their wavefunctions satisfy the condition
$|u_{\sigma}|=|v_{\sigma}|$. The contribution of the Majorana modes
to the order parameter is negligible. This is not surprising as many
states contribute to $\Delta(x)$.

\begin{figure}
\includegraphics[width=0.45\textwidth]{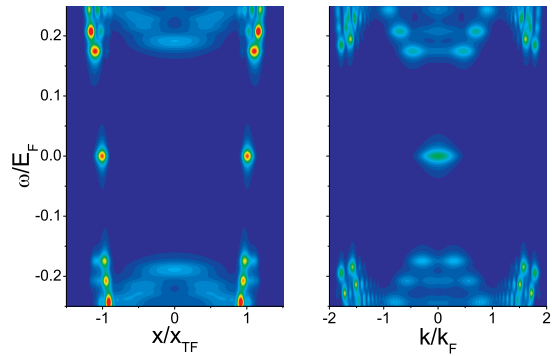}
\caption{Density of states in real space (left panel) and in momentum space
(right panel). The parameters are the same as in Fig.~\ref{f1}.
Brighter color represents higher density of states.}
\label{dos}
\end{figure}

Figure \ref{dos} shows the density of states both in real space and
momentum space and defined by
\begin{eqnarray*}
\rho(x,\omega) & = & \frac{1}{2}\underset{\sigma\eta}{\sum}[|u_{\sigma\eta}|^{2}\delta(\omega-E_{\eta})+|v_{\sigma\eta}|^{2}\delta(\omega+E_{\eta})]\,,\\
\tilde{\rho}(k,\omega) & = & \frac{1}{2}\underset{\sigma\eta}{\sum}[|\tilde{u}_{\sigma\eta}|^{2}\delta(\omega-E_{\eta})+|\tilde{v}_{\sigma\eta}|^{2}\delta(\omega+E_{\eta})]\,,
\end{eqnarray*}
 where $\tilde{u}_{\sigma\eta}(k)=\int u_{\sigma\eta}(x)e^{ikx}dx$,
$\tilde{v}_{\sigma\eta}(k)=\int v_{\sigma\eta}(x)e^{ikx}dx$ are the
Fourier transforms of $u_{\sigma\eta}(x)$ and $v_{\sigma\eta}(x)$,
respectively. In the calculations, the Dirac $\delta$-function is
replaced by a Gaussian with width on the order of spacings in the
energy spectrum away from the gap. The zero-energy Majorana modes
in the plots of the density of states are easily identified. They
are localized in both real space (near $\pm x_{TF}$) and momentum
space (around $k=0$). In principle, the density of states can be
measured in experiment by using spatial and momentum resolved radio-frequency
spectroscopy \cite{jiang11}.

We can derive analytic formulas for the Majorana wavefunctions using
the fact that they are localized in both real and momentum spaces.
The derivation with details found in the Supplemental Material \cite{sup} is
based on the linearization of the BdG Hamiltonian around $x=x_{0}$
and $k=0$, where $x_{0}$ is a real-space position near which the
Majorana modes are localized. A natural choice for $x_{0}$ is inspired
by the local density approximation at this point. The local, homogeneous
Hamiltonian becomes topologically nontrivial, has zero eigenvalues
when the Zeeman field strength $b=\sqrt{h^2+\nu^2}$ exceeds the critical value $\sqrt{\mu^{2}(x_{0})+|\Delta(x_{0})|^{2}}$,
where $\mu(x)=\mu-V_{T}(x)$. Hence we choose $x_{0}$ such that $b=\sqrt{\mu^{2}(x_{0})+|\Delta(x_{0})|^{2}}$.
Neglecting the kinetic energy term, approximating $V_{T}(x)$ by $\frac{1}{2}m\omega^{2}x_{0}^{2}+m\omega^{2}x_{0}(x-x_{0})$,
and replacing $\Delta(x)$ by $\Delta(x_{0})\equiv\Delta_{0}$ leads
to a linearized $H_{\rm BdG}$ . The zero energy modes of the linearized
$H_{\rm BdG}$ are found by using degenerate perturbation theory and leads
to Majorana modes that are Gaussians localized at either left or right
edge of the harmonic trap and given by
\[
|L,R\rangle=\frac{e^{-\frac{(x-x_{0})^{2}}{2\sigma^{2}}}}{\sqrt[4]{\pi\sigma^{2}}}\chi\;\;\mathrm{and}\;\;\sigma^{2}=\frac{\sqrt{|\Delta_{0}|^{2}-\nu^{2}}}{\sqrt{b^{2}-|\Delta_{0}|^{2}}}\frac{\lambda}{|m\omega^{2}x_{0}|}\;,
\]
where $|L,R\rangle$ are the left/right localized edge states, $\chi$
is a spatially independent Nambu function given in the Supplemental
Material \cite{sup}. We plot the analytical wavefunctions of the Majorana modes
in Fig.~\ref{f1}(d) together with the numerical results. The agreement
is remarkable.

\begin{figure}
\includegraphics[width=0.45\textwidth]{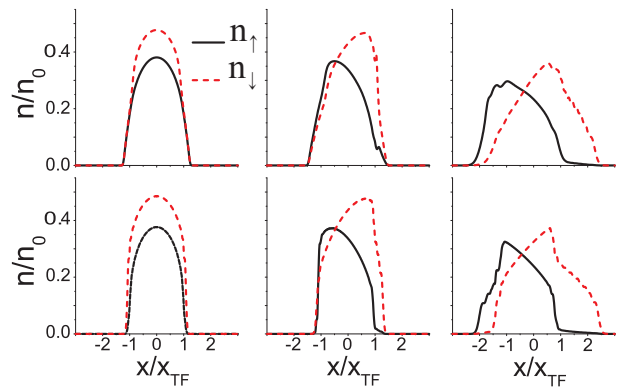}

\caption{Expansion dynamics of the Fermi cloud released at time $t=0$. Top
row: A interacting case with $\gamma=2.2$. Bottom row: A non-interacting
case with $\gamma=0$. Other parameters are: $\lambda=1.5E_{F}/k_{F}$,
$h=0.8E_{F}$ and $\nu=0.2E_{F}$. From left to right are plots of
density profiles at $t=0$, 0.88$/\omega_{0}$, and 2.04$/\omega_{0}$,
respectively.}

\label{exp}
\end{figure}

\begin{figure}
\includegraphics[width=0.4\textwidth]{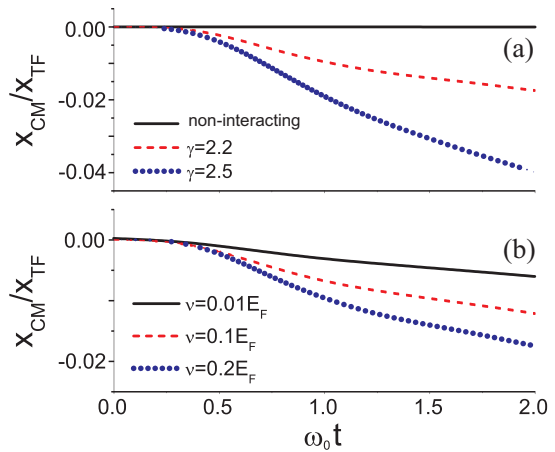}

\caption{Center-of-mass position of the atomic cloud during free expansion.
In (a), different curves represent different interaction strengths
for $\nu=0.2E_{F}$. In (b), different curves represent different
in-plane Zeeman fields for $\gamma=2.2$. Other parameters are the
same as in Fig.~\ref{exp}. }

\label{com}
\end{figure}

{\em Expansion of the FF state ---} The characteristic signature
of the FF state is finite-momentum pairing. Here, we show how this
feature manifests itself in the density profiles of the atomic cloud
during time-of-flight expansion. Figure~\ref{exp} shows the time
evolution of the density profiles for both an interacting and a non-interacting
gas. Initially, the density profiles of both spin states are symmetric
about the trap center. As the cloud expands, the profiles become asymmetric
when $\nu\neq0$. Furthermore, the center of mass of each spin state
moves in opposite directions regardless of whether the atoms interact
or not. There is, however, an important difference between an interacting
and a non-interacting cloud. As shown in Fig.~\ref{com}, for a non-interacting
cloud, the center-of-mass position of the whole cloud, defined as
\[
x_{{\rm cm}}=\frac{1}{N}\sum_{\sigma=\uparrow,\downarrow}\int x\, n_{\sigma}(x)\, dx\,,
\]
 remains at zero, while $x_{{\rm cm}}$ for an interacting cloud deviates
from zero as time increases. The deviation is stronger for larger
interaction strength $\gamma$ and larger in-plane Zeeman field strength
$|\nu|$. This result is consistent with a two-body calculation carried
out by Dong {\em et al.} \cite{dong13a}, where they found that
the total mechanical momentum of the interaction-induced two-body
bound state becomes finite as long as the in-plane Zeeman field is
present. In the present study, the non-zero $x_{{\rm cm}}$ during
the expansion is a direct consequence of the finite-momentum FF pairing
in the original trapped system. That $x_{{\rm cm}}$ changes faster
for larger interaction strength can be attributed to stronger pairing,
and hence a larger fraction of the atoms form Cooper pairs with finite
momentum.

{\em Conclusion ---} We have considered both the static and dynamical
properties of a trapped 1D Fermi gas subject to equal-weight Rashba-Dresselhaus
spin-orbit coupling. This system enters an exotic topological FF state
regime when a large effective Zeeman field with a non-zero in-plane
component is present. The two salient features of this phase is (1)
the presence of zero-energy Majorana modes localized near the edge
of the trap; and (2) finite-momentum pairing. These features manifest
themselves in the density of states and the center-or-mass displacement
during expansion, respectively. We hope future experiments may be
able to realize and probe this interesting quantum phase.

{\em Acknowledgements ---} LJ and ET acknowledge support from the
US Army Research Office under Contract No. 60661PH. XJL and HH are
supported by the ARC Discovery Projects (DP140100637, FT 130100815
and DP140103231). HP is supported by the NSF and the Welch Foundation
(Grant No. C-1669). We would like to thank Leslie Baksmaty for useful
discussions. 

\end{document}


\title{Supplemental Material}

\author{Lei Jiang$^{1}$, Eite Tiesinga$^{1}$, Xia-Ji Liu$^{2}$, Hui Hu$^{2}$
and Han Pu$^{3}$}

\affiliation{$^{1}$Joint Quantum Institute, University of Maryland and National
Institute of Standards and Technology, Gaithersburg, Maryland 20899,
USA\\
 $^{2}$Centre for Atom Optics and Ultrafast Spectroscopy, Swinburne
University of Technology, Melbourne 3122, Australia\\
 $^{3}$Department of Physics and Astronomy, and Rice Quantum Institute,
Rice University, Houston, Texas 77251, USA}

\date{\today }


\maketitle
In this Supplemental Material, we provide details on how to derive
the analytic wavefunction of the Majorana modes for a trapped 1D system.
Our method relies on a linearization of the Bogoliubov-de Gennes (BdG) Hamiltonian based on the
fact that Majorana modes are localized in both real (near the trap edges) and
momentum (near zero momentum) space. We divide the BdG Hamiltonian $H_{\rm BdG}=H_{0}+H_{1}+H_{2}$,
where $H_{0}$, a local Hamiltonian at real-space point $x_{0}$
near which the Majorana modes are localized, is given by
\[
H_{0}=\left(\begin{array}{cccc}
\frac{1}{2}m\omega^{2}x_{0}^{2}-\mu-h & -i\nu & 0 & -\Delta_{0}\\
i\nu & \frac{1}{2}m\omega^{2}x_{0}^{2}-\mu+h & \Delta_{0} & 0\\
0 & \Delta_{0}^{*} & -\frac{1}{2}m\omega^{2}x_{0}^{2}+\mu+h & -i\nu\\
-\Delta_{0}^{*} & 0 & i\nu & -\frac{1}{2}m\omega^{2}x_{0}^{2}+\mu-h
\end{array}\right)\,,
\]
 $H_{1}$ contains
the linearized potential  and spin-orbit coupling term
given by
\[
H_{1}=\left(\begin{array}{cccc}
m\omega^{2}x_{0}(x-x_{0}) & -i\lambda\hat{k} & 0 & 0\\
i\lambda\hat{k} & m\omega^{2}x_{0}(x-x_{0}) & 0 & 0\\
0 & 0 & -m\omega^{2}x_{0}(x-x_{0}) & i\lambda\hat{k}\\
0 & 0 & -i\lambda\hat{k} & -m\omega^{2}x_{0}(x-x_{0})
\end{array}\right)\,.
\]
Finally, $H_{2}$ contains the kinetic energy and corrections of
the order parameter $\Delta(x)$ near $\Delta_{0}\equiv\Delta(x_{0})$.
We neglect $H_{2}$ in our calculation.

The local Hamiltonian $H_{0}$ has two degenerate zero-energy states
when
\[
b=\sqrt{\left(\frac{1}{2}m\omega^{2}x_{0}^{2}-\mu\right)^{2}+|\Delta_{0}|^{2}}\,,
\]
where the effective Zeeman field strength $b=\sqrt{h^{2}+\nu^{2}}$
combines both the in- and out-of-plane field. This equation
sets the value of $x_{0}$. In our parameter region, $x_{0}$ has
two solutions $\pm|x_{0}|$, localized near the left and right edge of the
harmonic trap, respectively.
The corresponding zero-energy eigenstates  are
\begin{eqnarray*}
|1\rangle & = & \left(-\frac{a+h}{\sqrt{2b^{2}+2ah}},\,\frac{i\nu}{\sqrt{2b^{2}+2ah}},\,0,\,\frac{\Delta_{0}^{*}}{\sqrt{2b^{2}+2ah}}\right)^{T}\,,\\
|2\rangle & = & \left(\frac{i\nu}{\sqrt{2b^{2}-2ah}},\,\frac{a-h}{\sqrt{2b^{2}-2ah}},\,\frac{\Delta_{0}^{*}}{\sqrt{2b^{2}-2ah}},\,0\right)^{T}\,,
\end{eqnarray*}
where $a=\sqrt{b^{2}-|\Delta_{0}|^{2}}$.
The two other eigenstates of $H_0$ have finite energy $\pm2b$. As
we are only interested in zero-energy states of the  Hamiltonian,
we  neglect the effects of these two  energy states in the following.

Next, we treat $H_{1}$ as a small perturbation to $H_{0}$ in the subspace spanned by $\{|1\rangle,|2\rangle\}$. In this two dimensional subspace, $H_{1}$ takes the form
\[
H_{1}^{{\rm eff}}=\left(\begin{array}{cc}
\frac{a(a+h)}{b^{2}+ah}\tau\hat{x}-\frac{(a+h)\nu}{b^{2}+ah}\lambda\hat{k}, & -\frac{a\nu}{\sqrt{b^{4}-a^{2}h^{2}}}i\tau\hat{x}+\frac{\nu^{2}-|\Delta_{0}|^{2}}{\sqrt{b^{4}-a^{2}h^{2}}}i\lambda\hat{k}\\
\frac{a\nu}{\sqrt{b^{4}-a^{2}h^{2}}}i\tau\hat{x}-\frac{\nu^{2}-|\Delta_{0}|^{2}}{\sqrt{b^{4}-a^{2}h^{2}}}i\lambda\hat{k,} & \frac{a(a-h)}{b^{2}-ah}\tau\hat{x}-\frac{(a-h)\nu}{b^{2}-ah}\lambda\hat{k}
\end{array}\right)\,,
\]
where $\tau=m\omega^{2}x_{0}$ and $\hat{x}=x-x_{0}$.

In real space, the momentum operator takes the form $\hat{k}=-id/dx$ and we search for a zero-energy
eigenstate $(p,q)^{T}$ of $H_{1}^{{\rm eff}}$. This leads to
coupled differential equations for $p$ and $q$ with unit-normalized Gaussian
solutions
\begin{eqnarray*}
p & = & \frac{1}{\sqrt[4]{\pi\sigma^{2}}}e^{-\frac{(x-x_0)^{2}}{2\sigma^{2}}}\,,\\
q & = & -{\rm sgn}(\tau)\sqrt{\frac{h+a}{h-a}}\sqrt{\frac{b^{2}-ah}{b^{2}+ah}}\, p\,,
\end{eqnarray*}
where ${\rm sgn}(z)$ is the sign of $z$ and the width $\sigma$ is determined by $\sigma^{2}=\lambda\sqrt{h^{2}-a^{2}}/(|\tau|a)$.
The two corresponding normalized Nambu wavefunctions are
\[
|L,R\rangle = [u_{\uparrow }(x),u_{\downarrow }(x),v_{\uparrow }(x),v_{\downarrow }(x)]^{T} = \frac{1}{\sqrt[4]{\pi\sigma^{2}}}\,e^{-\frac{(x-x_{0})^{2}}{2\sigma^{2}}} \,\chi_{L,R} \,,\] with
\[\chi_{L,R} = \left(\frac{-\sqrt{h+a}[\sqrt{h^{2}-a^{2}}+{\rm sgn}(\tau)i\nu]}{2\sqrt{h}|\Delta_{0}|},\frac{\sqrt{h-a}[i\nu+{\rm sgn}(\tau)\sqrt{h^{2}-a^{2}}]}{2\sqrt{h}|\Delta_{0}|},-{\rm sgn}(\tau)\frac{\sqrt{h+a}}{2\sqrt{h}}\frac{\Delta_{0}^{*}}{|\Delta_{0}|},\frac{\sqrt{h-a}}{2\sqrt{h}}\frac{\Delta_{0}^{*}}{|\Delta_{0}|}\right)^{T}\,,
\]
where the left and right localized edge states $|L,R\rangle$ require ${\rm sgn}(\tau)=-1$ and 1, respectively.
One can check that the wavefunctions satisfy the symmetry requirement $|u_{\sigma}|=|v_{\sigma}|$
for Majorana modes. Finally, in order to compare with  numerical results
of our self-consistent BdG calculations, we define the symmetric and anti-symmetric states
$|\varphi^{\pm}\rangle=(|L\rangle\pm|R\rangle)/\sqrt{2}$.

In deriving the above results, we have assumed that $\nu<|\Delta_{0}|$.
This is reasonable as for large $\nu$ the order parameter becomes
vanishingly small. In the limit $\nu\ll|\Delta_{0}|$, the wavefunction
of the Majorana mode takes the form
\[
|L,R\rangle=\frac{1}{\sqrt[4]{\pi\sigma^{2}}}e^{-\frac{(x-x_{0})^{2}}{2\sigma^{2}}}\left(\frac{-(a+h)}{2\sqrt{h^{2}+ah}},\, {\rm sgn}(\tau)\frac{(h-a)}{2\sqrt{h^{2}-ah}},\,-{\rm sgn}(\tau)\frac{\Delta_{0}^{*}}{2\sqrt{h^{2}-ah}},\,\frac{\Delta_{0}^{*}}{2\sqrt{h^{2}+ah}}\right)^{T}\,.
\]